\documentclass{aastex63}
\usepackage{amssymb,amsmath}
\usepackage{graphicx}
\usepackage{xcolor,natbib}
\usepackage{multirow}
\usepackage[T1]{fontenc}
\usepackage{ae,aecompl}
\usepackage{newtxtext,newtxmath}
\usepackage{float}
\usepackage{subfigure}
\usepackage{longtable}
\usepackage{enumitem}
\usepackage{longtable,listings}
\usepackage[flushleft]{threeparttable}
\usepackage{parcolumns}
\submitjournal{ApJ}

\shorttitle{Dividing lines in BPT diagrams determined by t-SNE}
\shortauthors{Zhang et al.}

\begin{document}

\title{Powerful t-SNE technique leading to clear separation of type-2 AGN and HII galaxies in BPT diagrams}

\correspondingauthor{XueGuang Zhang}
\email{xgzhang@njnu.edu.cn}

\author{XueGuang Zhang}
\affiliation{School of Physics and technology, Nanjing Normal University,
          No. 1, Wenyuan Road, Nanjing, 210023, P. R. China}
\author{Yanqiu Feng}
\affiliation{School of Physics and technology, Nanjing Normal University,
          No. 1, Wenyuan Road, Nanjing, 210023, P. R. China}
\author{Huan Chen}
\affiliation{School of Physics and technology, Nanjing Normal University,
          No. 1, Wenyuan Road, Nanjing, 210023, P. R. China}
\author{QiRong Yuan}
\affiliation{School of Physics and technology, Nanjing Normal University,
          No. 1, Wenyuan Road, Nanjing, 210023, P. R. China}

\begin{abstract}
Narrow emission-line galaxies can be distinguished in the well-known BPT diagrams through narrow 
emission line properties. However, there are no boundaries visible to the naked eye between type-2 
AGN and HII galaxies in BPT diagrams, besides the extreme dividing lines expected by theoretical 
photoionization models. Here, based on powerful t-SNE technique applied to the local narrow 
emission-line galaxies in SDSS DR15, type-2 AGN and HII galaxies can be clearly separated in the 
t-SNE determined two-dimensional projected map, and then the dividing lines can be mathematically 
determined in BPT diagrams, leading to charming harmonization of the theoretical expectations and 
the actual results from real observed properties. The results not only provide an interesting and 
robust method to determine the dividing lines in BPT diagrams through the powerful t-SNE technique, 
but also lead to further confirmation on previously defined composite galaxies more efficiently 
classified in the BPT diagram of [OIII]/H$\beta$ versus [NII]/H$\alpha$.
\end{abstract}

\keywords{Galaxies:Emission line galaxies -- Galaxies:Galaxies nuclei -- Galaxies:AGN host galaxies}

\section{Introduction}

     BPT diagrams, named after \citet{bpt} and the pioneer work on detailed study on 
classifications of narrow emission-line galaxies in \citet{vo87}, are firstly proposed in 1980s 
to show different physical properties of optical narrow emission lines of tens to hundreds of 
extragalactic emission lines objects. Until now, there are millions of narrow emission-line galaxies. 
And based on different kinds of central activities, narrow emission-line galaxies can be well 
classified into two main kinds, type-2 AGN (Active Galactic Nuclei) (narrow emission-line 
AGN) with central AGN activities and HII galaxies without central AGN activities. Although Type-2 
AGN and HII galaxies have similar optical spectral features, properties of optical narrow emission 
lines can be commonly applied to determine classifications of type-2 AGN and HII galaxies, such 
as the well-known results in ongoing improved BPT diagrams \citep{csm11, jbc14, ks17, kn19}.

    The BPT diagrams well allow the division of narrow emission-line objects in two main branches: 
one containing HII galaxies and the other containing type-2 AGN (Seyfert galaxies and Low Ionization 
Nuclear Emission-Line Regions (LINERs)). Not similar as Seyfert galaxies totally powered by central 
AGN activities, different mechanisms have been applied to LINERs, such as AGN activities \citep{fn83, 
hs83}, shock heating \citep{ht80, ds95, ds96}, photoionization by young stars \citep{tm85, ft92}, 
photoionization by post-asymptotic giant branch (post-AGB) stars \citep{bm94, eh10, csm11}, etc. 
More recent review on LINERs in \citet{mm17} have shown that 60\% to 90\% of LINERs could be well 
considered as genuine AGN. Therefore, in spite of controversial mechanisms, LINERs have been accepted 
as a subsample of type-2 AGN in the manuscript.

   In the well-known BPT diagrams, between type-2 AGN and HII galaxies, there are no clear dividing 
boundaries visible to the naked eye. The reported dividing lines by flux ratios of optical narrow 
emission lines in the BPT diagrams are estimated and determined by expected properties of extreme 
starbursts and/or AGNs by theoretical photoionization models, such as the results well discussed 
in \citet{kh03, gd04, kg06, sc06, lk10, mh14}, etc. More recently, \citet{ds17} have 
studied emission-line galaxy classifications through probabilistic Gaussian mixture model applied 
to spectroscopic properties from the SDSS DR7 (Sloan Digital Sky Survey, Data Release 7) and 
SEAGal/STARLIGHT datasets, and shown the Gaussian components relative to AGN and starforming galaxies. 
However, the results discussed in \citet{ds17} can not yet lead to apparent dividing lines. Thereby, 
it is a great pity on loss of determining the dividing lines from properties of observed narrow emission 
lines of real narrow emission-line galaxies. Here, based on the more recent powerful t-SNE (t-distributed 
Stochastic Neighbor Embedding) technique \citep{vh08, vm14, ah18} which have been recently applied in 
Astrophysics \citep{tm17, ac18, sw20}, we will show actual results on the dividing lines in BPT diagrams 
from real observed properties of narrow emission lines of local narrow emission-line galaxies, to check 
whether there are harmonization of theoretical expectations and actual results on the dividing lines 
between HII galaxies and type-2 AGN in the BPT diagrams. Then, our main results and necessary discussions 
are shown in Section 2, and conclusions are given in Section 3.

\begin{figure*}
\centering\includegraphics[width = 18cm,height=6.5cm]{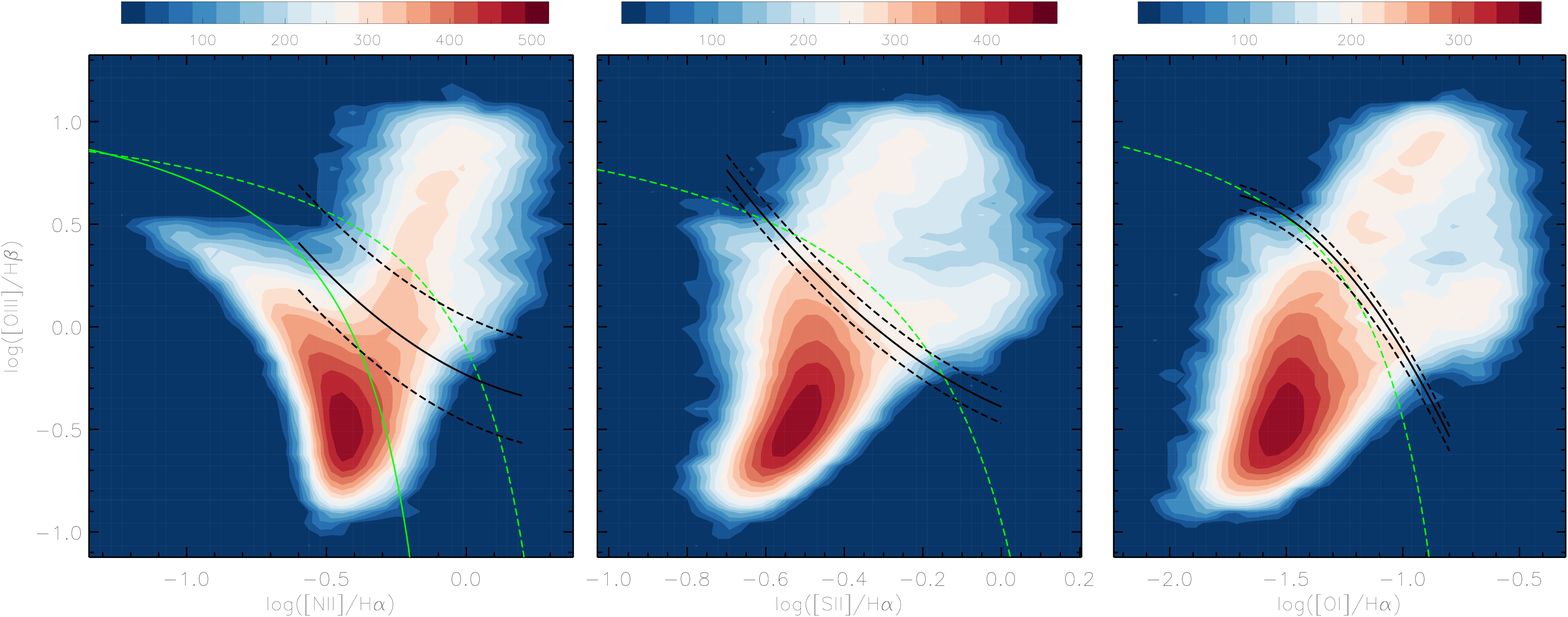}
\centering\includegraphics[width = 18cm,height=6.5cm]{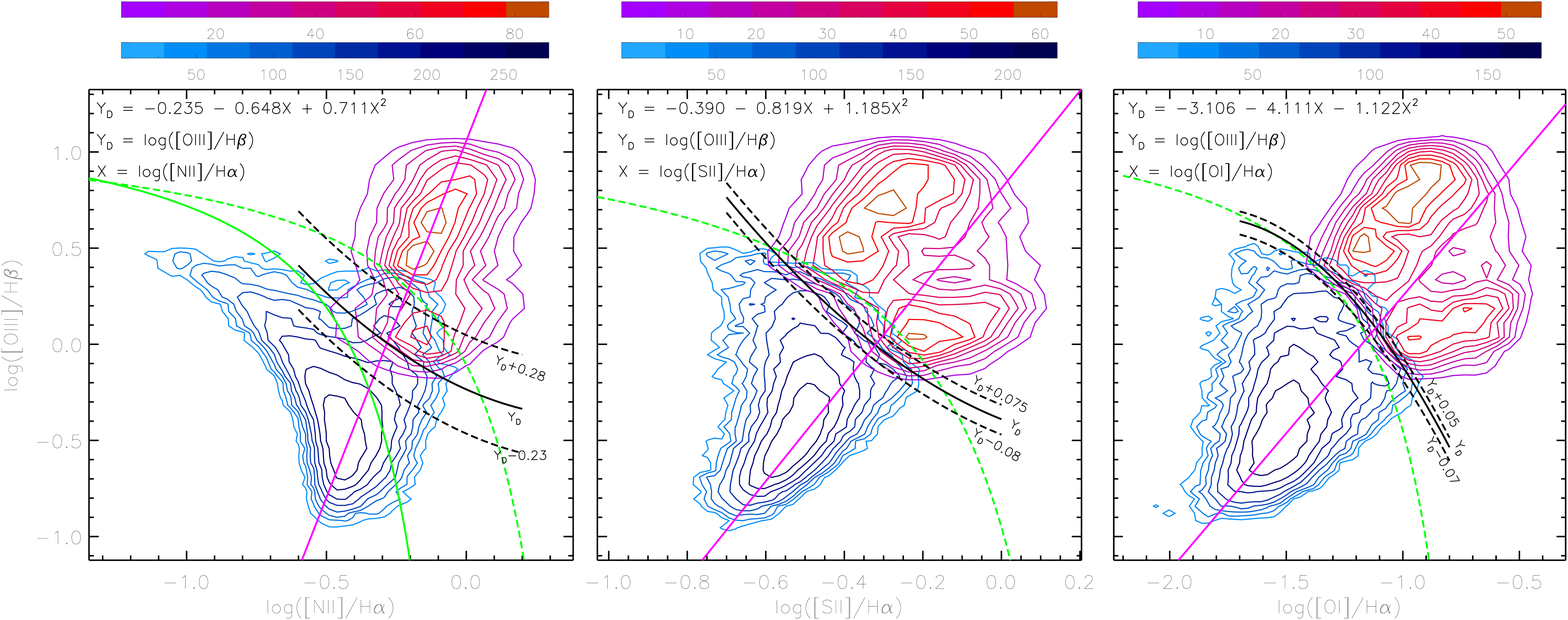}
\caption{Properties of local narrow emission-line galaxies and dividing lines in BPT diagrams. 
{\bf{top panels}} show the well-known three BPT diagrams in contour maps by flux ratios of [OIII]/H$\beta$, 
[NII]/H$\alpha$, [SII]/H$\alpha$ and [OI]/H$\alpha$ for the collected 35857 local narrow emission-line 
galaxies in SDSS DR15. Color bars show the corresponding number densities for the contour levels. 
{\bf{Bottom panels}} show the corresponding two contour maps in the three BPT diagrams for the objects 
in the two clusters shown in Figure~\ref{tsne} determined by the t-SNE technique: the contour with 
contour lines near to red colors represents the map for the 7169 objects shown in red dots in 
Figure~\ref{tsne}, the contour with contour lines near to blue colors represents the map for the 28688 
objects shown in blue dots in Figure~\ref{tsne}. Color bars with colors near to blue and near to red 
show the corresponding number densities for the levels in the two contours, respectively. In each bottom 
panel, solid line in magenta shows the referenced direction, in order to create series of strips to 
determine the crossover data points of the position dependent number ratios of the objects in the two 
clusters shown in Figure~\ref{tsne} included in the strips. In each panel, dashed line in green shows 
the photoionization model determined dividing line between extremely starburst galaxies and type-2 AGN 
(or the extreme outer boundary for HII galaxies) reported in Kewley et al (2001, 2006). 
Solid lines in green in the left panels show the dividing line between HII galaxies and composite 
galaxies reported in Kauffmann et al. (2003). In each panel, solid line in black shows 
the t-SNE technique determined dividing line $Y_D$, the expected dividing lines between type-2 AGN and 
HII galaxies, and dashed lines in black show the corresponding upper and lower boundaries of 
$Y_{\rm upper}$ and $Y_{\rm lower}$.
}
\label{bpt}
\end{figure*}

\section{Main Results and Discussions}

    Based on properties of optical narrow emission lines of main galaxies, the SQL query 
(\url{http://skyserver.sdss.org/dr15/en/tools/search/sql.aspx}) can be conveniently applied in 
SDSS DR15 \citep{aa19, bw19}, leading to collected 35857 local narrow emission line objects with 
apparent narrow emission lines (line intensities at least 5 times larger than their corresponding 
measured errors) of H$\alpha$, H$\beta$, [O~{\sc iii}]$\lambda$5007\AA, [O~{\sc i}]$\lambda$6300\AA, 
[N~{\sc ii}]$\lambda$6583\AA~ and [S~{\sc ii}]$\lambda$6717, 6731\AA~ but no broad emission lines in 
high-quality SDSS spectra (signal-to-noise ratios larger than 10). The applied SQL query in detail 
is as follows,
\begin{lstlisting}
SELECT plate, fiberid, mjd, z, snmedian, h_beta_flux, h_beta_flux_err, h_alpha_flux, 
       h_alpha_flux_err, oiii_5007_flux, oiii_5007_flux_err, nii_6584_flux, 
       nii_6584_flux_err, sii_6717_flux, sii_6717_flux_err, sii_6731_flux, 
       sii_6731_flux_err, oi_6300_flux, oi_6300_flux_err 
FROM GalSpecLine as G JOIN SpecObjall as S ON S.specobjid = G.specobjid
WHERE S.class = 'GALAXY' and S.SNmedian > 10 and S.z < 0.2 and S.zwarning = 0
       and G.h_beta_flux_err > 0 and G.h_beta_flux > 5*G.h_beta_flux_err and 
       G.h_alpha_flux_err > 0 and G.h_alpha_flux > 5*G.h_alpha_flux_err and 
       G.oiii_5007_flux_err > 0 and G.oiii_5007_flux > 5* G.oiii_5007_flux_err and 
       G.nii_6584_flux_err > 0 and G.nii_6584_flux > 5*G.nii_6584_flux_err and 
       G.oi_6300_flux_err > 0 and G.oi_6300_flux > 5* G.oi_6300_flux_err and 
       G.sii_6717_flux_err > 0 and G.sii_6717_flux > 5* G.sii_6717_flux_err and 
       G.sii_6731_flux_err > 0 and G.sii_6731_flux > 5* G.sii_6731_flux_err and 
       (S.subclass = 'STARFORMING' or S.subclass = 'STARBURST' or S.subclass = 'AGN') 
       and G.h_alpha_flux/G.H_beta_flux < 6 and G.h_alpha_flux/G.H_beta_flux > 2 
       and veldisp >60 and veldisp < 400 and veldisperr > 0 and veldisp > 5*veldisperr
\end{lstlisting}.

   Then, based on the collected properties of narrow emission lines of the 35857 local narrow 
emission-line galaxies, beautiful BPT diagrams are shown in top panels in Figure~\ref{bpt}, It is 
clear there are no apparent boundaries visible to the naked eye between the expected type-2 AGN 
locating in top-right regions and the expected HII galaxies locating in bottom-left regions in the 
shown BPT diagrams. Therefore, it is a constructive challenge to determine the dividing lines 
through mathematical visualization techniques applied to the real observed narrow emission-line 
galaxies, besides the theoretical model determined ones. 

\begin{figure*}
\centering\includegraphics[width = 12cm,height=10cm]{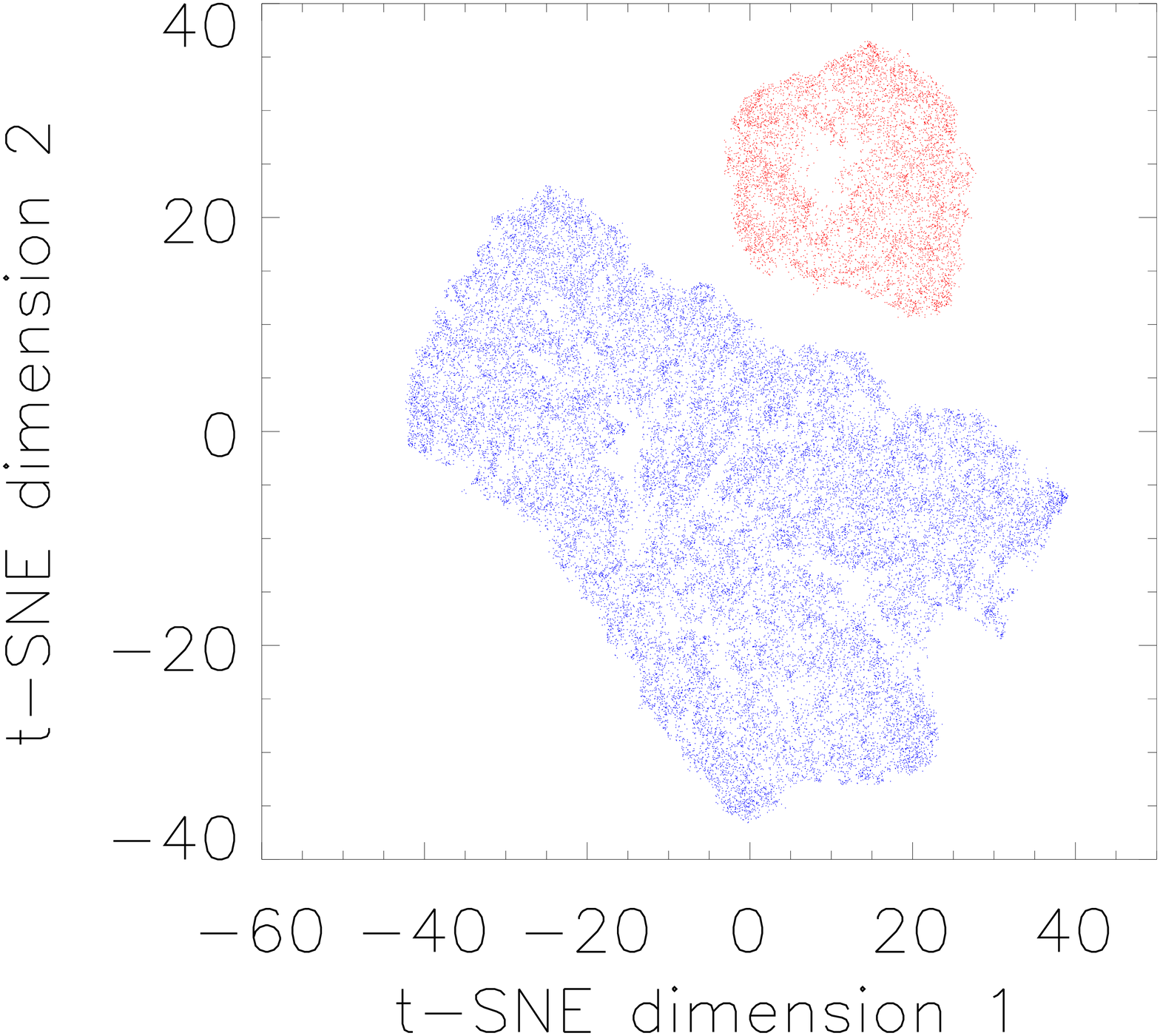}
\caption{The t-SNE technique determined 2D projection of the collected 35857 local narrow 
emission-line galaxies through properties of the four narrow line ratios of [OIII]/H$\beta$, 
[NII]/H$\alpha$, [SII]/H$\alpha$ and [OI]/H$\alpha$ applied in the BPT diagrams. There are 
two clusters: one cluster with dots in red and the other cluster with dots in blue.}
\label{tsne}
\end{figure*}

    The two well-known mathematical techniques have been accepted to do visualization and reduction 
of high dimensional data, the commonly known PCA (Principal Component Analysis) technique \citep{jc16, 
lk17} and the more recent powerful t-SNE technique \citep{vh08, vm14, ah18}. The PCA 
technique has been well applied in Astronomy for several decades, such as the results reported in 
\citet{sc97, bc98, wh05, rb07, sm09}.The main idea behind the PCA technique through deterministic 
algorithm without hyperparameters is linear technique to reduce the dimensionality of data that is 
highly correlated by transforming the original set of vectors to a new set known as Principal 
Components to preserve the global structure of the data. In the manuscript, the first two most 
important Principal Components (PCA dimension 1 and PCA dimension 2) are held by rotating the vectors 
for preserving variance. Meanwhile, the more recent powerful t-SNE technique involving hyperparameters 
is a non-linear technique through non-deterministic or randomised algorithm. The math behind t-SNE 
is quite complex but the idea is simple. It embeds the points from a higher dimension to a lower 
dimension trying to preserve the neighborhood of that point, preserving the local structure (cluster) 
of data. More specifically, t-SNE technique minimizes the divergence between two distributions: 
a distribution that measures pairwise similarities of the high-dimensional data and a distribution 
that measures pairwise similarities of the corresponding low-dimensional points in the embedding. 
In the manuscript, the determined similarity of data points are shown in the corresponding reduced 
two-dimensional embedded space, i.e., the t-SNE dimension 1 and t-SNE dimension 2. Moreover, the 
commonly accepted hyperparameters applied in the t-SNE technique are perplexity, early exaggeration, 
learning rate and number of steps (the hyperparameters listed in sklearn.manifold.TSNE in python). 
Not similar as PCA technique, the t-SNE technique cannot preserve variance (global structure) but 
preserve distance using hyperparameters (local structure). Therefore, the non-linear and 
probabilistic t-SNE technique can often find fine structures where the PCA technique cannot.

    Considering the four-dimensional data including the line ratios of [O~{\sc iii}]$\lambda$5007\AA~ 
to narrow H$\beta$ ([OIII]/H$\beta$), [N~{\sc ii}]$\lambda$6583\AA~ to narrow H$\alpha$ ([NII]/H$\alpha$), 
total [S~{\sc ii}] to narrow H$\alpha$ ([SII]/H$\alpha$) and [O~{\sc i}]$\lambda$6300\AA~ to narrow 
H$\alpha$ ([OI]/H$\alpha$) of the 35857 local narrow emission-line objects, the four line ratios 
applied in the three BPT diagrams, the data reduction and visualization can be well done through 
the two mathematical techniques. Figure~\ref{tsne} shows the two-dimensional projected map through 
the t-SNE technique applied to the four-dimensional data of the 35857 narrow emission-line objects. 
Here, the accepted hyperparameters applied in the t-SNE technique are [170,~5,~400,~800] for perplexity, 
early exaggeration, learning rate and number of steps, respectively. As we know that 
the powerful t-SNE technique is a particularly better technique than the PCA technique to especially 
do visualization of high-dimensional datasets, the t-SNE technique can clearly lead to nice two 
clusters shown in blue dots for 28688 objects and in red dots for 7169 objects in Figure~\ref{tsne}. 
Meanwhile, the PCA technique cannot lead to similar results, compared results from the PCA technique 
and from the t-SNE technique are shown in the following Figure~\ref{tsne2} and Figure~\ref{tsne3}.

    The nice two clusters well determined by the t-SNE technique can strongly indicate intrinsic 
different physical properties in the BPT diagrams for the objects in the two clusters shown 
in Figure~\ref{tsne}. Then, the narrow emission-line objects in the two clusters determined by the 
t-SNE technique are separately replotted in the BPT diagrams in bottom panels of Figure~\ref{bpt} 
with contour lines in bluish colors (for the objects shown in blue dots in Figure~\ref{tsne}) and 
in reddish colors (for the objects shown in red dots in Figure~\ref{tsne}), respectively. Then, 
based on the contours with contour lines in different colors in the BPT diagrams, it is interesting 
that the narrow emission-line objects in the two clusters shown in Figure~\ref{tsne} can be well 
distinguished in the BPT diagrams in bottom panels of Figure~\ref{bpt}, providing the chance to 
determine dividing lines between HII galaxies and type-2 AGN by independent mathematical methods.

\begin{figure*}
\centering\includegraphics[width = 18cm,height=13cm]{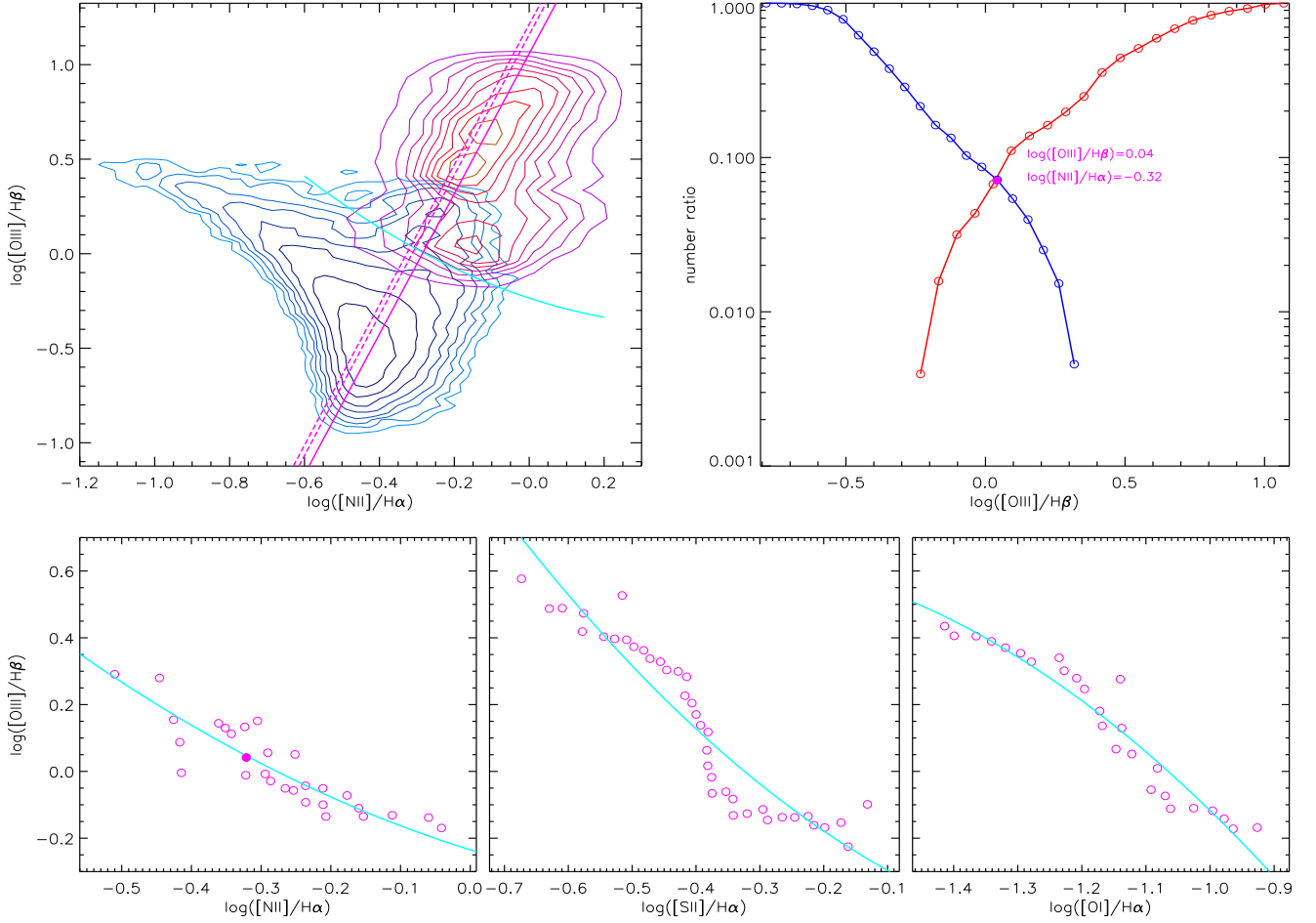}
\caption{Detailed example on determining dividing lines in BPT diagrams. {\bf Top-left panel} 
shows an example of the strip marked by two dashed lines in magenta with directions parallel 
to the given referenced direction shown in the solid line in magenta (same as the one shown 
in bottom-left panel of Figure~\ref{bpt}). Solid line in cyan shows the determined dividing line. 
The two contours represent the maps for the objects in the two clusters shown in Figure~\ref{tsne}. 
{\bf Top-right panel} shows the normalized cumulative position-dependent number ratios of the 
objects included in the strip shown in the top-left panel from the objects in the two clusters 
with contour maps shown in colors near to blue and in colors near to red. Solid circle in magenta 
in the panel shows the crossover point of the two position dependent number ratios of the objects 
from the two clusters in the strip shown in the top-left panel. The position information of the 
crossover point is marked with characters in magenta. {\bf Bottom panels} show the determined 
crossover points shown as open circles in magenta from a series of strips in the three BPT diagrams, 
and the 2-degree polynomial functions shown as solid lines in cyan applied to describe the data 
points. Solid circle in magenta in the bottom-left panel shows the crossover point determined 
in the top-right panel by the objects in the strip shown in the top-left panel.}
\label{dline}
\end{figure*}

    In the BPT diagram of [OIII]/H$\beta$ versus [NII]/H$\alpha$, there are two well accepted 
dividing lines between HII galaxies, composite galaxies and type-2 AGN, one is applied to determine 
the outer boundary for extremely starburst galaxies, and the other one is applied to determine the 
pure HII galaxies without central AGN activities, the region between the two dividing lines are 
well defined for composite galaxies. Now, in the bottom-left panel of Figure~\ref{bpt}, the two 
previously defined dividing lines can be roughly compared with properties of the objects in the 
two clusters, considering the far-side outer boundaries for the objects in the two clusters. 
Meanwhile, the theoretical model determined dividing lines in the other two BPT diagrams can also 
be well compared with almost overlapped far-side boundaries of the objects in the two clusters 
determined by the t-SNE technique.

    In order to construct a definite dividing line in the BPT diagram of [OIII]/H$\beta$ versus 
[NII]/H$\alpha$ between the objects in the two clusters determined by the t-SNE technique, a simple 
method has been applied. Followed a series of parallel strips with directions parallel to the 
referenced direction linked the ridges of the two contour maps of the objects in the two clusters, 
the dividing line $Y_D$ is well built by the crossover points of the position dependent number 
ratios for the objects in the two clusters in the strips. The final dividing line $Y_D$ has been 
well determined and shown in bottom-left panel of Figure~\ref{bpt}. In the other two BPT diagrams, 
the referenced directions are given by the direction linked the ridge of the contour map in bluish 
colors and the gully of the other contour map in reddish colors, then leading to the t-SNE technique 
determined dividing lines $Y_D$. Moreover, totally similar results can be determined and confirmed 
in the BPT diagrams by given slightly different referenced directions but followed from the 
bottom-left to the top-right in the BPT diagrams. Here, Figure~\ref{dline} shows an example on 
more detailed information on the crossover points for the objects in the two clusters in strips 
in the BPT diagrams.

\begin{figure*}
\centering\includegraphics[width = 18cm,height=6.5cm]{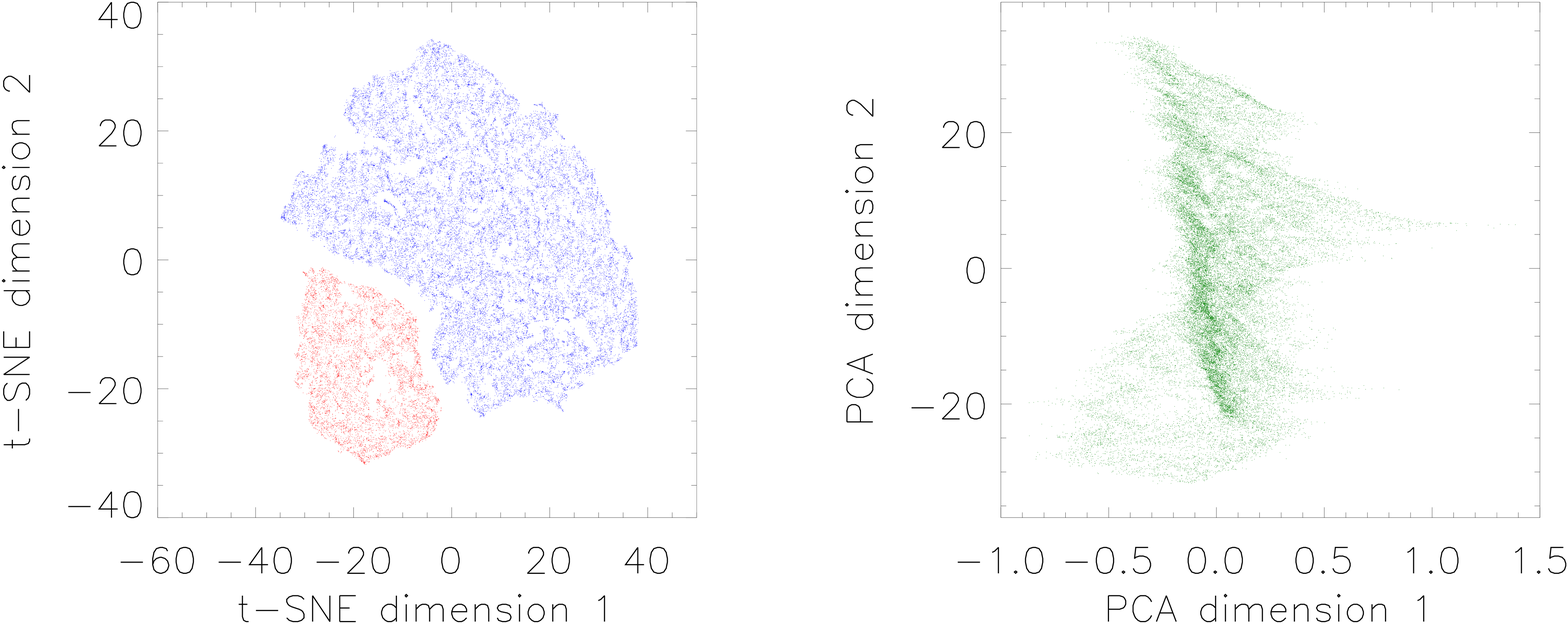}
\centering\includegraphics[width = 18cm,height=6.5cm]{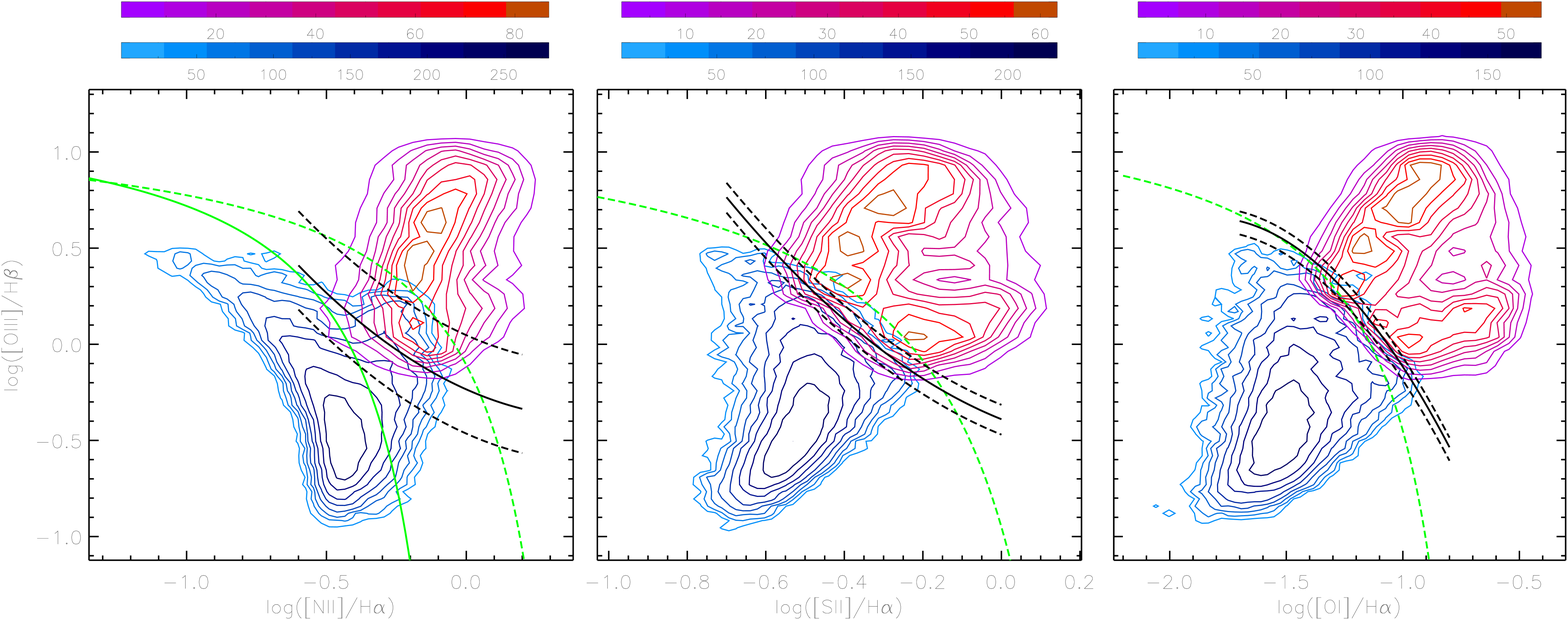}
\caption{Properties of narrow emission-line galaxies and dividing lines in BPT diagrams by t-SNE
technique with different hyper-parameters. {\bf{Top panels}} show the 2D projections of the 
four-dimensional data of narrow emission line ratios of the 35857 narrow emission-line galaxies 
through the t-SNE technique (top-left panel) with hyper-parameters different from the ones applied 
in Figure~\ref{tsne} and through the PCA technique (top-right panel). {\bf{Bottom panels}} show 
the corresponding two contour maps in the three well-known BPT diagrams for the objects in the 
two clusters determined by the t-SNE technique: the contour with contour lines near to red colors 
represents the map for the objects shown in red dots in the top-left panel, and the contour with 
contour lines near to blue colors represents the map for the objects shown in blue dots in the 
top-left panel. Color bars in different colors show the corresponding number densities for the 
contour levels. In bottom panels, dashed lines in green show the photoionization model determined 
dividing lines between extremely starburst galaxies and type-2 AGN reported in 
Kewley et al. (2001, 2006). Solid line in green in the bottom-left panel shows the dividing line 
between HII galaxies and composite galaxies reported in Kauffmann et al. (2003). 
Solid and dashed lines in black in each bottom panel show the dividing line $Y_D$ and the 
corresponding upper and lower boundaries of $Y_{\rm upper}$ and $Y_{\rm lower}$, the ones shown 
in Figure~\ref{bpt}.}
\label{tsne2}
\end{figure*}

\begin{figure*}
\centering\includegraphics[width = 18cm,height=6.5cm]{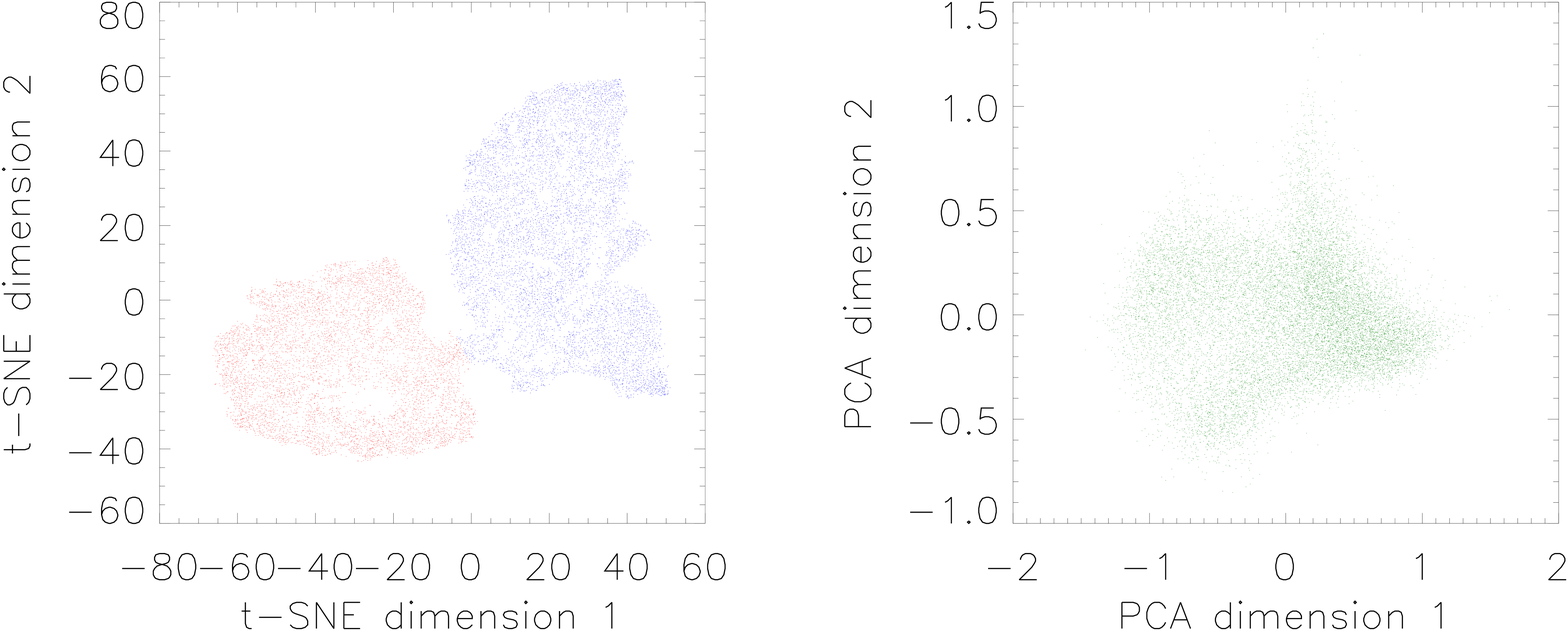}
\centering\includegraphics[width = 18cm,height=6.5cm]{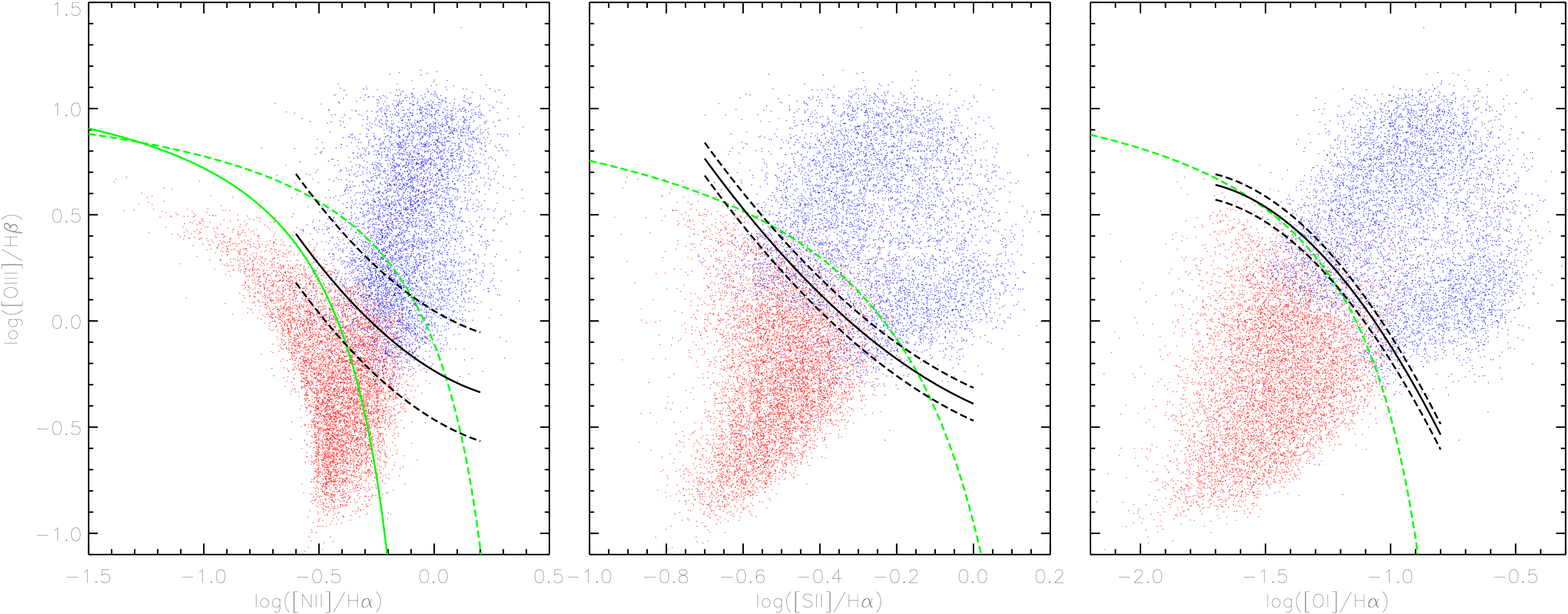}
\caption{Properties of narrow emission-line galaxies and dividing lines in BPT diagrams by the t-SNE
technique applied to the smoothed sample of narrow emission-line galaxies. {\bf{Top panels}} show the
2D projections of the four-dimensional data of narrow emission line ratios of the smoothed 14554 narrow
emission-line galaxies through the t-SNE technique (top-left panel) and through the PCA technique
(top-right panel). {\bf{Bottom panels}} show the three BPT diagrams for the 14554 galaxies in the
two clusters determined by the t-SNE technique: small pluses in red and blue represent the objects 
shown in red dots and in blue dots in the top-left panel. In bottom panels, dashed lines in green 
show the photoionization model determined dividing lines between extremely starburst galaxies and 
type-2 AGN reported in Kewley et al. (2001, 2006). Solid line in green in the bottom-left 
panel shows the dividing line between HII galaxies and composite galaxies reported in 
Kauffmann et al. (2003). Solid and dashed lines in black in each bottom panel show the dividing line 
$Y_D$ and the corresponding upper and lower boundaries of $Y_{\rm upper}$ and $Y_{\rm lower}$, the 
ones shown in Figure~\ref{bpt}.}
\label{tsne3}
\end{figure*}

    Meanwhile, besides the t-SNE technique determined dividing lines $Y_D$ described by the two-degree 
polynomial functions shown in Figure~\ref{bpt} and Figure~\ref{dline}, it will be very interesting to 
discuss properties of overlapped regions covered by both HII galaxies and type-2 AGN in the BPT diagrams. 
It is clear that when the objects in the two clusters are plotted in the BPT diagrams with 
their measured flux ratios of narrow emission lines (such as the results shown in bottom panels of 
Figure~\ref{bpt}), there are overlapped regions in the BPT diagrams for the objects in the two clusters 
determined by the t-SNE technique. Therefore, the overlapped regions in the BPT diagrams are very real, 
which will provide further clues on properties of composite galaxies. Then, a simple mathematical idea 
is accepted to determine the overlapped regions between $Y_{\rm upper}~=~Y_D~+~U$ and 
$Y_{\rm lower}~=~Y_D~-~B$. Here, parameters of $U$ and $B$ are determined by the criteria that 1\% of 
HII galaxies are above $Y_{\rm upper}$ and 1\% of type-2 AGN are below $Y_{\rm lower}$ in the BPT 
diagrams of [NII]/H$\alpha$ versus [OIII]/H$\beta$ and [SII]/H$\alpha$ versus [OIII]/H$\beta$, but 0.5\% 
of HII galaxies are above $Y_{\rm upper}$ and 0.5\% of type-2 AGN are below $Y_{\rm lower}$ in the 
BPT diagram of [OI]/H$\alpha$ versus [OIII]/H$\beta$ (mainly due to the following shown very smaller 
overlapped region). The well determined boundaries $Y_{\rm upper}$ and $Y_{\rm lower}$ are shown in 
Figure~\ref{bpt}.

     For the results shown in bottom-left panel of Figure~\ref{bpt}, the overlapped region is 
determined between $Y_{\rm upper}~=~Y_D~+~0.28$ and $Y_{\rm lower}~=~Y_D~-~0.23$ in the BPT diagram 
of [NII]/H$\alpha$ versus [OIII]/H$\beta$, meaning that 99\% of HII galaxies are locating below the 
upper boundary of $Y_{\rm upper}$ and 99\% of type-2 AGN are locating above the lower boundary of 
$Y_{\rm lower}$. Therefore, we can simply say that the overlapped regions defined by the upper and 
lower boundaries of $Y_{\rm upper}$ and $Y_{\rm lower}$ represent the mixed region of HII galaxies 
and type-2 AGN with significant confidence level of 99\%. Furthermore, the region between $Y_{\rm upper}$ 
and $Y_{\rm lower}$ has interestingly similar covered space as the region between the reported 
dividing lines in \citet{ke01, kg06} and in \citet{kh03}. However, there are no reports on composite 
galaxies in the BPT diagrams of [SII]/H$\alpha$ versus [OIII]/H$\beta$ and [OI]/H$\alpha$ versus 
[OIII]/H$\beta$ in the literature. We directly accepted the same criterion applied in the BPT diagram 
of [NII]/H$\alpha$ versus [OIII]/H$\beta$. Then, the upper and lower boundaries of 
$Y_{\rm upper}~=~Y_D~+~0.075$ and $Y_{\rm lower}~=~Y_D~-~0.08$ have been determined and shown in the 
bottom-middle panel of Figure~\ref{bpt} in the BPT diagram of [SII]/H$\alpha$ versus [OIII]/H$\beta$. 
Meanwhile, the overlapped region by the same criterion in the BPT diagram of [OI]/H$\alpha$ 
versus [OIII]/H$\beta$ is too small to be shown. Thereby, a smaller percentage of 0.5\% than 1\% is 
applied to determine the upper and lower boundaries of $Y_{\rm upper}~=~Y_D~+~0.05$ and 
$Y_{\rm lower}~=~Y_D~-~0.07$ shown in the BPT diagram of [OI]/H$\alpha$ versus [OIII]/H$\beta$ in 
the bottom-right panel of Figure~\ref{bpt}.

    It is very interesting that there is a larger overlapped area between $Y_{\rm upper}$ and 
$Y_{\rm lower}$ for composite galaxies in the BPT diagram of [NII]/H$\alpha$ versus [OIII]/H$\beta$, 
but much smaller overlapped areas in the BPT diagrams of [OI]/H$\alpha$ versus [OIII]/H$\beta$ and 
[SII]/H$\alpha$ versus [OIII]/H$\beta$. Therefore, the BPT diagram of [NII]/H$\alpha$ versus 
[OIII]/H$\beta$ is more efficient and powerful to classify the composite galaxies than the other 
two BPT diagrams. And moreover, the defined composite galaxies in the BPT diagram of [NII]/H$\alpha$ 
versus [OIII]/H$\beta$ cannot be totally in the right places for composite galaxies in the other two 
BPT diagrams of [OI]/H$\alpha$ versus [OIII]/H$\beta$ and [SII]/H$\alpha$ versus [OIII]/H$\beta$. The 
different properties of composite galaxies in different BPT diagrams reflect interesting but 
different dependence of forbidden emission lines on intrinsic different kinds of activities from 
AGN or from star-forming.

   Finally, more detailed discussions have been shown on robust results through the t-SNE technique 
by the following two ways. On the one hand, besides the results shown in Figure~\ref{bpt} and 
Figure~\ref{tsne}, Figure~\ref{tsne2} shows similar results confirmed through the t-SNE technique applied 
with different hyper-parameters to the same four-dimensional data of narrow line ratios of the collected 
35857 narrow emission-line galaxies. Here, the accepted hyperparameters re-applied in the 
t-SNE technique are [50,~10,~80,~1000] for perplexity, early exaggeration, learning rate and number of 
steps, respectively. Two definite data clusters can be re-confirmed and similar final results can 
be re-confirmed in the BPT diagrams as the results shown in Figure~\ref{bpt}. On the other hand, in 
order to reduce computational complexity, especially to reduce probable effects of much different 
number densities of the objects in different regions of BPT diagrams, the BPT diagrams have been well 
smoothed leading to the almost even number densities in different regions with total 14554 objects 
collected from the 35857 objects, which are shown in bottom panels of Figure~\ref{tsne3}. Then, the 
similar t-SNE and PCA techniques are applied to do the data reduction and visualization of the 
four-dimensional data of the smoothed 14554 narrow emission-line galaxies. Here, the 
accepted hyperparameters applied in the t-SNE technique are [150,~5,~100,~800] for perplexity, early 
exaggeration, learning rate and number of steps, respectively. Top panels of Figure~\ref{tsne3} 
show the 2D projections by the t-SNE and PCA techniques. Nice two clusters can be well confirmed 
by the t-SNE technique. Bottom panels of Figure~\ref{tsne3} show properties of the objects in the 
two clusters and then the dividing lines applied in the BPT diagrams for the 14554 narrow 
emission-line galaxies.

   Before the end of the section, there is one point we should note. As what we have known, there are 
two subsamples included in type-2 AGN: Seyfert galaxies and LINERs, and moreover, there are reported 
dividing lines between LINERs and Seyfert galaxies in the BPT diagrams, such as the results shown in 
\citet{kg06} and in \citet{st07}. However, as the discussed results in \citet{ds17}, the Gaussian 
mixture model can not provide statistical evidence for the existence of a Seyfert/LINER dichotomy for 
the emission-line galaxies from SDSS DR7. Moreover, as the shown results through the t-SNE technique in 
Figure~\ref{tsne}, in top-left panel of Figure~\ref{tsne2} and in top-left panel of Figure~\ref{tsne3}, 
we cannot yet find a third data cluster determined by the t-SNE technique. More efforts should be necessary 
on constructing a new high-dimensional data including more valuable information. Therefore, at the 
current stage, we can not provide clear clues on the dividing lines between LINERs and Seyfert galaxies 
through the t-SNE technique, and we do not show further discussions any more on dividing lines between 
LINERs and Seyfert galaxies in the manuscript.

\section{conclusions}

    BPT diagrams are powerful tools for classifying narrow emission-line galaxies with different central 
activity properties. However, how to define the dividing lines in the BPT diagrams between different kinds 
of narrow emission-line galaxies is always an interesting challenge. Here, we can find well defined 
dividing lines through the pure mathematical t-SNE technique applied to the local narrow emission-line 
galaxies in SDSS DR15. The results not only show the charming harmonization of the theoretical expectations 
and the actual results from real observed properties through the powerful t-SNE technique, and also provide 
further confirmation on classification of the composite galaxies more efficiently in the BPT diagram of 
[OIII]/H$\beta$ versus [NII]/H$\alpha$.

\section*{Acknowledgements}
{\color{red}Zhang, Feng, Chen and Yuan gratefully acknowledge the anonymous referee for 
giving us constructive comments and suggestions to greatly improve our paper.} 
Zhang gratefully thanks the grant support from Nanjing Normal University and the grant 
support from NSFC-11973029. Yuan gratefully thanks the grant support from NSFC-11873032. 
This paper has made use of the data from the SDSS projects. The SDSS-III web site is 
http://www.sdss3.org/. SDSS-III is managed by the Astrophysical Research Consortium for 
the Participating Institutions of the SDSS-III Collaboration.

\end{document}